\documentclass[
 amsmath, amssymb, breqn,
 aps, prl, superscriptaddress, twocolumn,
longbibliography]{revtex4-1}

\usepackage{graphicx}
\usepackage{amsmath,amssymb}
\usepackage{textcomp}

\newcommand{\bv}[1]{\ensuremath{\mathbf{#1}}}

\def\3He{$^3$He}
\def\4He{$^4$He}

\begin{document}


\title{Surface-dominated finite size effects in nanoconfined superfluid helium}

\author{E. Varga}
\email{ev@ualberta.ca}
\author{C. Undershute}%
\author{J. P. Davis}
\email{jdavis@ualberta.ca}
\affiliation{%
 Department of Physics, University of Alberta, Edmonton, Alberta T6G 2E1, Canada
}

\date{\today}

\begin{abstract}
Superfluid $^4$He (He-II) is a widely studied model system for exploring finite-size effects in strongly confined geometries. Here, we study He-II confined in mm-scale channels of 25 and 50 nm height at high pressures using a nanofluidic Helmholtz resonator. We find that the superfluid density is measurably suppressed in the confined geometry from the transition temperature down to 0.6 K. Importantly, this suppression can be accounted for by roton-like thermal excitation with an energy gap of 5 K. We show that the surface-bound excitations lead to the previously unexplained lack of finite-size scaling of suppression of the superfluid density.
\end{abstract}

\maketitle



Physical descriptions of condensed matter typically implicitly assume infinite sizes of the studied samples. When surfaces, finite sizes, or restricted geometries are considered, however, novel behaviors often emerge. These range widely across the field of condensed matter, e.g., the quantum hall effect in 2D electron gas \cite{Stormer1999,Hansson2017,vonKlitzing2020}, conducting surface states in topological insulators \cite{Hasan2010}, or the topological phase transition in the 2D XY model \cite{Kosterlitz1972,Kosterlitz1973}. Finite-size effects play a significant role in, e.g., technologically-relevant quantum-dots \cite{Lodahl2015} or properties of nanoparticles \cite{Pankhurst2003}, where magnetic phase transitions can be strongly altered or even completely suppressed compared to the bulk material \cite{Kubaniova2019} or novel collective behaviors can emerge \cite{Halperin1986,Batlle2002}. Similarly, in superfluid \3He, surface scattering of Cooper pairs can lead to stabilisation of novel superfluid phases \cite{Shook2020,Yapa2022}.

Finite-size effects become especially important near phase transitions, where the coherence length diverges \cite{Gasparini2008}. The superfluid phase transition in \4He is one of the most closely studied model systems thanks to the high achievable purity of the medium and relative lack of parasitic effects due shape or interaction with confining walls \cite{Barmatz2007}. Indeed, the most accurate experimental determination of a critical exponent occurred in \4He in a microgravity environment \cite{Barmatz2007}. A detailed understanding of the superfluid transition in \4He is not only useful for tests of modern phase transition theories \cite{Barmatz2007}, but also for systems in the same universality class \cite{Nishimori2010}.

Due to the ability of superfluid helium (He-II) to easily flow in most strongly confined systems, the superfluid transition is a popular model system for the study of finite size effects near the phase transition. The superfluid order parameter is the macroscopic wave function $\Psi$, related to the superfluid density as $\rho_s = |\Psi|^2$ \cite{Tilley_book}, which vanishes at the walls \cite{Lea1989}. This results in suppression of superfluid density in confined geometries, which, for 2D crossover, is expected \cite{Gasparini2008} to follow
\begin{equation}\label{eq:finite-size-scaling}
    \frac{\rho_{sc}}{\rho} = \frac{\rho_{sb}}{\rho}\left( 1 - X(lt^\nu)\right),
\end{equation}
where $X$ is a universal function, $\rho$ is the total density, $\rho_{sc}$ and $\rho_{sb}$ are the superfluid densities in the confined geometry and bulk, respectively, $t = 1 - T/T_\lambda$ is the reduced temperature ($T_\lambda$ is the bulk transition temperature), $l = D/\xi_0$ is the reduced system size with $D$ the thickness of the slab, $\xi_0$ the low-temperature coherence length, and $\nu\approx 0.67$ is the correlation length critical exponent.

For He-II in the 2D-limit, as the critical temperature $T_c$ is approached, the superfluid density vanishes discontinuously \cite{Nelson1977} at the Kosterlitz-Thouless (KT) transition \cite{Berezinski1971,Kosterlitz1972,Kosterlitz1973}. The KT transition was tested to a high degree of accuracy \cite{Reppy1992}, however, the scaling behavior given by \eqref{eq:finite-size-scaling}, is not observed experimentally and the reason for the breakdown is not known \cite{Kimball2001,Gasparini2001,Gasparini2008,Gasparini2021}. Breakdown of finite size scaling was also observed in thermal resistivity \cite{Murphy2003}. This is in contrast with good agreement of finite-size scaling of the specific heat above $T_\lambda$ \cite{Murphy2003,Kimball2004,Kimball2005}. Note that the scaling near the phase transition in 3D and 2D is connected via hyperscaling relations \cite{Barmatz2007}, anomalous behavior in He-II thus requires close attention. 

\begin{figure*}
    \centering
    \includegraphics{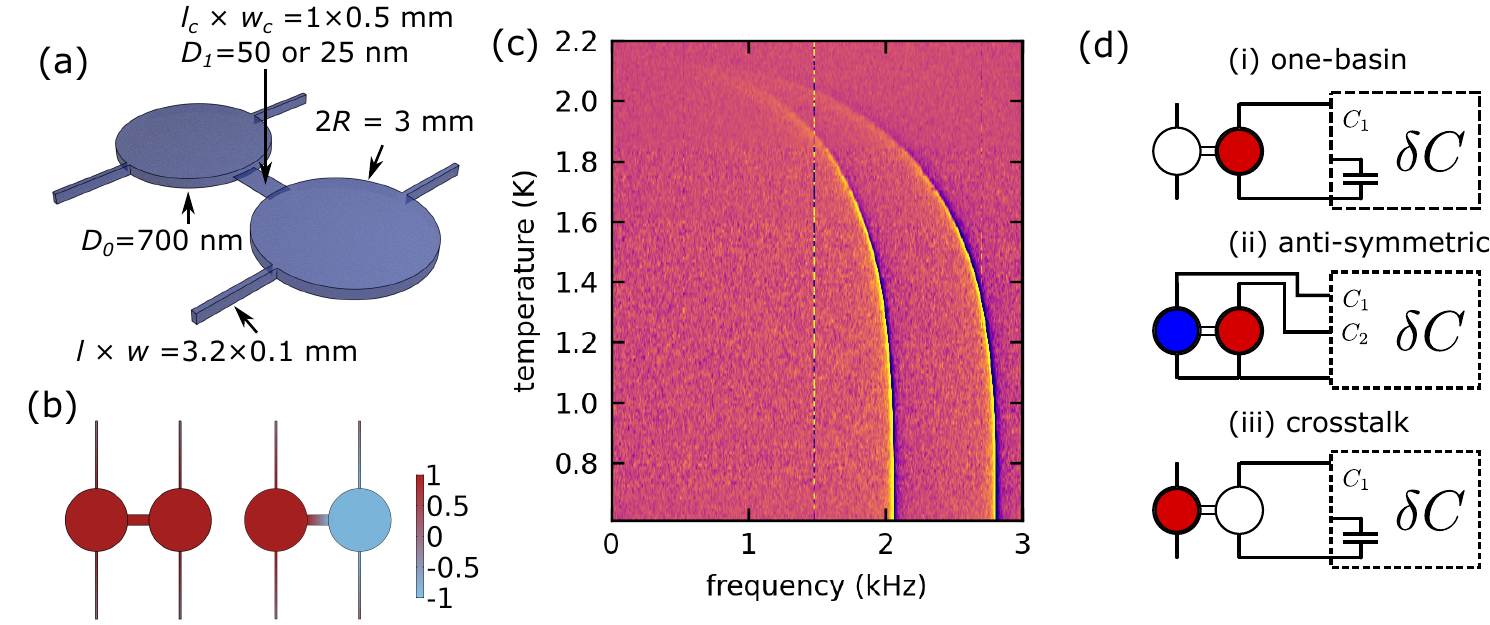}
    \caption{(a) The confined volume of the nanofluidic Helmholtz resonator. The two circular basins are connected to a pressurized \4He bath via four inlet channels and are interconnected via the strongly confined central channel (see Supplementary Material for AFM images). The basins form parallel plate capacitors (electrodes not shown) used for both sensing and forcing the Helmholtz modes. (b) Pressure amplitude of the two Helmholtz modes used in the study. Note that the fundamental (symmetric) mode shows zero pressure gradient across the strongly confined channel making only the second (anti-symmetric) mode sensitive to the superfluid density in the strong confinement. (c) Temperature dependence of the two modes for 50 nm central channel confinement and $P\approx 1$ bar, measured with the one-basin protocol. (d) The three measurement protocols (see Supplementary Material for more detailed description). The ``$\delta C$" box represents a bridge circuit that measures differential capacitance between one of the basins and a reference capacitor (the one-basin (i) and cross-talk (iii)) or between the two basins (the anti-symmetric (ii)). The coloring of the basin indicates the applied electrostatic force driving the Helmholtz mode: white -- not forced, red -- forced, blue -- forced with a 180\textdegree~phase shift.}
    \label{fig:setup}
\end{figure*}

Here, we observe the suppression of superfluid density in fully confined helium slabs with 25 nm and 50 nm confinements probed using an on-chip nanofluidic Helmholtz resonance, where tuning the bulk excitation spectrum, coherence length, and the transition temperature is possible by changing the pressure. We observe the suppression for temperatures 0.6 K -- $T_\lambda$ due to roton-like excitation with energy gap of 5 K (independent of confinement or pressure) that is localized near the walls. We find that the suppression of the superfluid density at different confinements and pressures, which, notably, does not follow finite-size scaling \eqref{eq:finite-size-scaling}, can be fully accounted for by pressure and confinement dependence of the roton wave-vector. The helium slabs exhibit the KT transition, which indicates that the coherence length reaches a significant fraction of the confinement close to the transition \cite{Kosterlitz1972}. In particular, the nature of the superfluid suppression does not change between high-temperature, the 2D regime, and the low-temperature 3D regime where the coherence length is much shorter than the slab thickness, showing that the surface excitations, rather than coherence length effects, are the dominant process behind finite size effects in the studied range of parameters. This resolves the long-standing puzzle of breakdown of finite-size scaling in superfluid \4He.

We measure the superfluid fraction $\rho_s/\rho$ using the 4th sound resonance method \cite{Rojas2015,Thomson2016,Souris2017,Shook2020}. The 4th sound resonance is set up in a nanofluidic Helmholtz resonator, Fig.~\ref{fig:setup}(a,b), where He-II is confined to a thin volume enclosed by a quartz substrate (see \cite{Souris2017} for fabrication details). Here, the resonator differs from previous designs \cite{Varga2020,Shook2020} by using two separate circular volumes (``basins", confinement $D_0\approx 700$~nm) interconnected through a strongly confined central channel (confinement $D_1 \approx $ 25 or 50~nm). The nanofluidic volume is connected to a surrounding pressurised bath via four inlets. This geometry supports two Helmholtz modes, whose pressure amplitude is shown in Fig.~\ref{fig:setup}(b), which differ in the relative phase of the pressure oscillating in the two basins. In the fundamental mode (frequency $\omega_0$), the pressure oscillates in the two basins in phase, whereas in the second Helmholtz mode (frequency $\omega_1$) the pressure in the basins oscillates with a 180\textdegree~phase shift. The temperature dependence of the two modes is shown in Fig.~\ref{fig:setup}(c), where the color indicates the the magnitude of one quadrature of the response.

Normalizing by the zero-temperature frequencies $\omega_0(0)$ and $\omega_1(0)$, we get for the superfluid fraction of the bulk (see Supplementary Material for derivation)
\begin{equation}
    \label{eq:frac-rhos-bulk}
    \frac{\rho_{sb}}{\rho} = \frac{\omega_0^2(T)}{\omega_0^2(0)},
\end{equation}
and for the confined channel
\begin{equation}
    \label{eq:frac-rhos-conf}
    \frac{\rho_{sc}}{\rho} = \frac{\omega_1^2(T) - \omega_0^2(T)}{\omega_1^2(0) - \omega_0^2(0)}.
\end{equation}
The above expressions neglect the temperature dependence of the total density, which varies by 0.6\% from 0 K to $T_\lambda$ \cite{Donnelly1998}. Experimentally, the lowest attainable temperature was 0.6 K, which was used in place of the zero-temperature limit (for bulk He-II at saturated vapour pressure, $\rho_s/\rho > 99.99\%$ at 0.6 K \cite{Donnelly1998}).

Since no measurable deviations from bulk behavior are expected for the 700 nm confinement further than 1 mK from the transition temperature \cite{Gasparini2008,Rojas2015}, we use the superfluid fraction determined using the fundamental mode as an in-situ thermometer calibrated against the bulk superfluid fraction calculated using the HEPAK data \cite{HEPAK}. We drive both resonances sufficiently weakly to avoid turbulent non-linear response \cite{Varga2020}. 

The helium motion is driven and detected using parallel plate capacitors deposited in the device basins, which are wired in a tuned capacitance bridge circuit allowing sensitive detection of pressure fluctuations inside the basin. We check the robustness of the measured superfluid densities using three different experimental protocols shown in Fig.~\ref{fig:setup}(d). With the one-basin protocol (i), both modes are excited and detected using only one basin and the two modes show no relative phase shift. With the anti-symmetric protocol (ii), only the second Helmholtz mode is driven and observable below the critical temperature $T_c$ of the confined channel. Above $T_c$, the two basins become decoupled and the fundamental mode frequency is observed, since the mode is degenerate with respect to the phase of the pressure oscillation in the two basins. Finally, the cross-talk protocol (iii) separates the drive and detection. Below $T_c$ we observe both modes with approximately 180\textdegree~phase shift and above $T_c$ no response is observed (see Supplementary Material for details).

\begin{figure}
    \centering
    \includegraphics{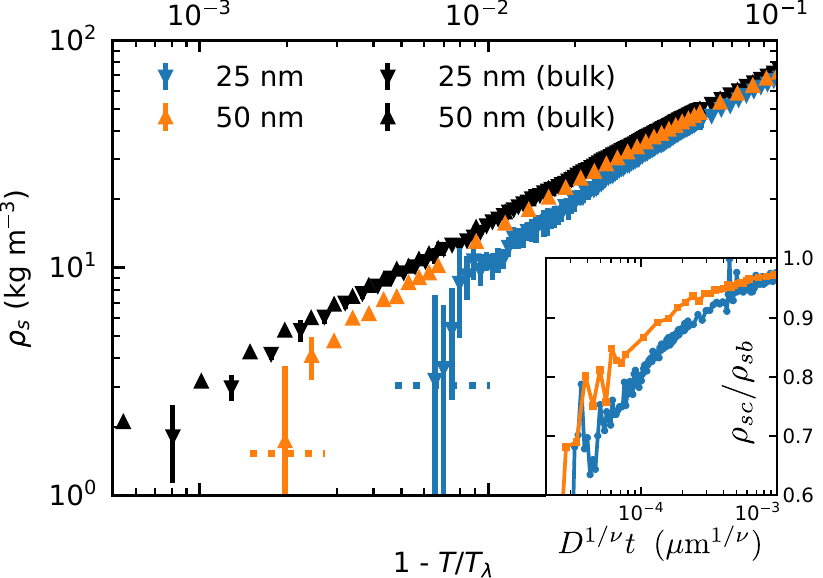}
    \caption{Superfluid density as a function of reduced temperature in the transition region. The confinement of the central channel of the resonator is shown in the legend. The confined superfluid density is calculated using \eqref{eq:frac-rhos-conf} and the black points show the bulk superfluid density for the corresponding device calculated using \eqref{eq:frac-rhos-bulk}. Thick dotted horizontal lines indicate the expected KT jump \cite{Nelson1977}. The error bars are calculated as the standard deviation of a set of measurements with different protocols (i-iii) and different drive amplitudes. Insert: ratio of confined and bulk superfluid densities as a function of the scaling variable according to \eqref{eq:finite-size-scaling}.  Note that our data does not follow a universal scaling, as has been previously observed \cite{Gasparini2008}.}
    \label{fig:KT}
\end{figure}

The temperature dependence of the superfluid density in the transition region for the 25 and 50 nm confinements at approximately 1 bar is shown in Fig.~\ref{fig:KT}. Shown by dotted horizontal lines are the expected universal jumps for the two confinements in superfluid density at the KT transition \cite{Nelson1977}, which agree well with our data. The universal jump has been experimentally verified in helium films \cite{Bishop1978,Minnhagen1987a}, and to lesser degree also in fully confined geometries \cite{Gasparini2008,Perron2010,Perron2012,Thomson2016}.

\begin{figure*}
    \centering
    \includegraphics{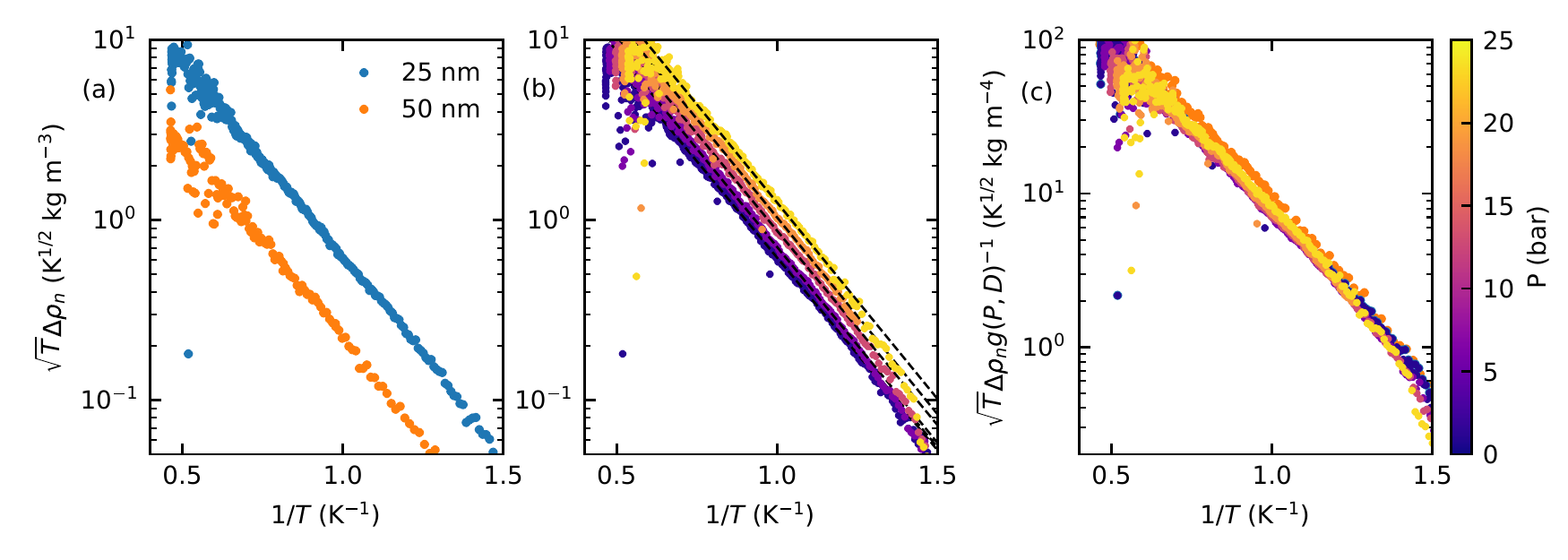}
    \caption{(a) Normal fluid density enhancement for helium slabs with 25~nm and 50~nm confinements. (b) Temperature dependence of the normal fluid density enhancement as a function of pressure for 25 nm~confinement. The black dashed lines in (a,b) are linear fits to \eqref{eq:roton-rhon}, the roton gap is given by the slope of the line. (c) Data from panels (a) and (b) scaled with $g(P,D) = D^{-1}(1 + P/P^*)(1 + (D^*/D)^3)$ with $P^*=25$~bar and $D^*=25$~nm (see text).}
    \label{fig:rhon-enhancement}
\end{figure*}

The ratio of the confined and bulk superfluid densities is shown in the insert of Fig.~\ref{fig:KT} as a function of scaling variable $D^{1/\nu}t$. According to the scaling relation \eqref{eq:finite-size-scaling}, the data should fall on a universal locus. As in previous experiments \cite{Gasparini2008}, this is not the case.

Instead, we interpret the suppression of superfluid density as an enhancement of normal fluid density $\rho_n = \rho - \rho_s$ via surface-bound thermal excitations. The density $\rho_n$ is determined by the spectrum of thermal excitations $\varepsilon(k)$ \cite{StatisticalPhysics1969}, with the dominant contribution above approximately 1 K due to the roton minimum, where
\begin{equation}
    \label{eq:roton-energy}
    \varepsilon^\mathrm{rot}(k) = \Delta + \frac{\hbar^2(k-k_0)^2}{2m^*}.
\end{equation}
Here $\Delta/k_B \approx 9$~K, $k_0 \approx 2$~\AA$^{-1}$, and $m^*\approx 0.14m_4$, with $m_4$ the mass of \4He atom, for bulk helium \cite{Godfrin2021}. The resulting roton contribution to the normal fluid density in two dimensions is \cite{StatisticalPhysics1969,Chester1976} (in units of mass per area)
\begin{equation}
\label{eq:roton-rhon} 
    \rho_n^\mathrm{2D}(T) = \frac{\hbar k_0^3 m^{*1/2}}{2\sqrt{2\pi k_B T}}\exp\left(-\frac{\Delta}{k_B T}\right).
\end{equation}
The normal fluid density enhancement, $\Delta\rho_n = \rho_{sc} - \rho_{sb}$, is shown in in Fig.~\ref{fig:rhon-enhancement}(a) for the two confinements and pressure $P\approx 1$ bar and in Fig.~\ref{fig:rhon-enhancement}(b) for several $P\in[1, 25]$~bar for the 25 nm confinement as a function of inverse temperature. Plots in Fig.~\ref{fig:rhon-enhancement}(a,b) are scaled in a manner such that a relation of the type of Eq.~\eqref{eq:roton-rhon} appears as a straight line with the slope equal to $-\Delta/k_B$. The temperature dependence in Figs.~\ref{fig:rhon-enhancement}(a,b) shows excellent agreement with \eqref{eq:roton-rhon} in the full temperature range. Despite the fact that the shown data cover more than an order of magnitude in $\xi(T)/D$, no significant deviation from \eqref{eq:roton-rhon} is observed, which demonstrates that the coherence length does not play a significant role in the observed suppression of the superfluid density.

The roton gap is found to be $\Delta\approx5$~K, independent of temperature and pressure (see Supplementary Material for additional details), which is in agreement with previous studies with porous materials and helium films, where a temperature dependence of normal fluid density consistent with a roton gap of $\Delta/k_B \approx 4 - 6$ K was observed \cite{Kiewiet1975,Chester1976,Bishop1981}. Excitations below the bulk roton gap were also observed via inelastic neutron scattering \cite{Clements1993,Clements1996}. Numerically, it was shown in 2D and quasi-2D helium layers that the roton gap should be significantly reduced compared to the bulk value \cite{Padmore1974}, and it was suggested that this was due to sharper structure function peak in 2D liquids and backflow enhancement around the roton core \cite{Gotze1976}. It is also believed that 2D roton excitations exist in the few helium monolayers adjacent to the solidified layers on the substrate \cite{Chester1980,Bishop1981} and only a weak dependence of the reduced gap on the area density (i.e., thickness) is predicted \cite{Apaja2003,Arrigoni2013}. However, the pressure-independent roton gap is in contrast to past observations in bulk \cite{Godfrin2012} and porous media \cite{Yamamoto2008,Bossy2012}, suggesting that the behavior of disordered 3D surfaces is more complex.

Adopting the standard bulk-plus-surface approach \cite{Gasparini2008}, we assume that the 2D rotons are localized to a thin layer of dense liquid adjacent to the walls. The resulting enhancement of the measured 3D normal fluid density in the confined channel is
\begin{equation} \label{eq:delta-rhon}
    \Delta\rho_n (T) = \frac{2}{D} \rho_n^\mathrm{2D}(T).
\end{equation}
While the measured normal fluid enhancements for two different confinements in Fig.~\ref{fig:rhon-enhancement}(a) clearly differ only by a scaling constant, \eqref{eq:delta-rhon} predicts scaling $\Delta\rho_n\propto D^{-1}$ which does not collapse the data. Similar to the effect of the confinement, a change in pressure only re-scales the normal fluid enhancement by a temperature-independent constant, as shown in Fig.~\ref{fig:rhon-enhancement}(b) (see Supplementary Material for pressure dependence of the pre-factor in \eqref{eq:roton-rhon}). Numerical calculations have shown \cite{Arrigoni2013} that the 2D roton wave-vector $k_0$ depends on the layer density approximately linearly (we neglect the weak variation of the gap with density \cite{Apaja2003,Arrigoni2013}). Assuming that the density near the wall changes linearly with bulk pressure we can write $k_0(P) = k_P' + k_P'' P$. Retaining only the first order in $P$, the normal fluid enhancement should scale as $\Delta\rho_n\propto (1 + P/P^*)$, which is found to collapse the data well with $P^*\approx 25$~bar. In this light, confinement dependence can be interpreted as a decrease in density in the boundary region as the separation between the walls increases. Indeed, the pressure at distance $z$ from the wall, outside of the first few solidified layers,  varies approximately as $P(z) = P + \alpha/z^3$, where $P$ is the bulk pressure and $\alpha$ depends on the substrate \cite{Chester1976,Brooks1979}. This results in $k_0(D) \approx k_D' + k_D'' D^{-3}$ and $\Delta\rho_n \propto (1 + (D^*/D)^3)D^{-1}$, which does collapse the data for $D^*\approx 25$~nm. The overall confinement and pressure dependence can thus be described as $\Delta\rho_n\propto g(P,D)$ with $g(P,D) = D^{-1}(1 + P/P^*)(1 + (D^*/D)^3)$, which is shown in Fig.~\ref{fig:rhon-enhancement}(c). Since only two confinements are presently available, different relations could equally well describe the data, such as $k_0\propto D^{-1/6}$ or many others. The scaling with $g(P,D)$ is appealing, however, since it describes pressure and confinement dependence in a unified way, as variation of the fluid density close to the wall.

The exact understanding of the roton behavior in confined slab geometry (e.g., calculation of $D^*$ and $P^*$) will require numerical calculations of the excitation spectrum for confinements in the range of tens of nm, which are within reach of present-day computational resources \cite{Apaja2001,Apaja2003a,Krotscheck2007}. Nevertheless, the important observation stemming from the present data is that the suppression of superfluidity in 2D confined geometries is dominated by surface excitations, specifically 2D rotons, rather than coherence-length effects as was previously assumed \cite{Gasparini2008}. Even a relatively weak dependence of the roton wave vector on the confinement is a natural explanation of the breakdown of finite-size scaling in confined thin slabs of superfluid \4He. Interestingly, since the growth of the coherence length near $T_\lambda$ is terminated by the KT transition (or the smallest dimension of the system itself), it is questionable whether a regime of coherence length scaling is ever obtained in real systems. Finally, we also note that the breakdown of scaling due to surface effects was considered also in Ref.~\cite{Gasparini2001}, although the 2D roton was not identified and the inclusion of the wall attraction on $T_\lambda$ leads to even greater discrepancy with the experiment \cite{Wang1990,Mooney2002}.

In conclusion, using a nanofluidic Helmholtz resonator capable of in-situ comparison of bulk-like and strongly confined behavior of superfluid helium we have shown that the dominant mechanism of suppression of the superfluid density in planar confined geometry are roton-like quasiparticles with an energy gap of 5 K localized near the wall. This appears to be the dominant mechanism of $\rho_s$ suppression for both 25 and 50 nm confinements for a wide range of pressure and temperature, up-to only a few mK below $T_\lambda$. A relatively weak dependence of the roton minimum on the confinement naturally explains the lack of coherence-length scaling which was a long-standing unresolved puzzle \cite{Gasparini2008}.

Finite size effects in confined superfluid helium remain an area of active research \cite{Gasparini2008,Perron2010,Kulchytskyy2013,DelMaestro2017,Gasparini2021} due, in part, to the ability of modern nanofabrication methods to confine liquid helium to precisely engineered geometries. The general assumption that the confining walls provide simply a termination for the macroscopic wave function \cite{Gasparini2008} is shown to be incomplete. Surface-bound excitations have strong effect on the dynamics of the confined superfluid, which will likely play a role in future precision tests of modern phase transition theories \cite{Barmatz2007} or novel quantum applications of thin helium layers \cite{He2020,Sfendla2021}.

\begin{acknowledgments}
This work was supported by the University of Alberta; the Natural Sciences and Engineering Research Council, Canada (Grant Nos.~RGPIN-2016-04523 and RGPIN-2022-03078); and the Alberta Quantum Major Innovation Fund. C.U.~acknowledges support from the University of Alberta Department of Physics SUPRE program.
\end{acknowledgments}

\onecolumngrid
\appendix
\newpage
\section*{Supplementary Material}

\renewcommand{\figurename}{Fig.~}
\renewcommand\thefigure{SM~\arabic{figure}}
\renewcommand\theequation{S\arabic{equation}}
\setcounter{figure}{0}
\setcounter{equation}{0}

\section{Resonance frequencies of a two-basin Helmholtz resonator}

The derivation of the resonance frequencies of the multiple Helmholtz modes follows along similar lines to Refs.~\cite{Souris2017,Varga2020,Varga2021} with an additional degree of freedom due to the motion of the fluid inside the central channel.

Let $y_{1,2}$ denote the displacement of helium in the inlet channels connected to basin 1 and 2 ($y_{1,2}>0$ for inflow into the basin, $y_{1,2}<0$ for outflow) and $y_c$ the displacement of helium in the central channel ($y_c > 0$ being flow from basin 1 to basin 2). Similarly, let $x_{1,2}$ be the displacement of the quartz plates of the basins with $x_{1,2} > 0$ being expansion.

As the superfluid component of helium flows through the channels, the total mass in the basins will change as
\begin{subequations}
    \label{eq:delta-mass}
    \begin{align}
        \delta m_1 &= 2\rho_{sb} a y_1 - \rho_{sc}a_c y_c \\
        \delta m_2 &= 2\rho_{sb} a y_2 + \rho_{sc}a_c y_c,
    \end{align}
\end{subequations}
where $\rho_{sb}$ is the bulk superfluid density and $\rho_{sc}$ is the superfluid density in the strongly confined channel. $a$ is the cross section of the inlet channel and $a_c$ is the cross section of the central channel. The total density of liquid in the basins changes by 
\begin{equation}
    \label{eq:delta-rho}
    \delta\rho_{1,2}=\delta\left(\frac{m_{1,2}}{V_{B1,2}}\right) = \frac{\delta m_{1,2}}{V_B} - 2\frac{\rho A x_{1,2}}{V_B},
\end{equation}
where $V_B = AD$ is the volume of the basin of area $A=\pi R^2$ and height $D$, and $\rho$ is the total density. The change of liquid density in the basin is related to the change in pressure through compressibility $\chi$ as $\delta p_{1,2} = \delta\rho_{1,2}/(\rho\chi)$. Neglecting inertia of the quartz plates themselves (admissible as long as the their fundamental resonance frequency is much higher than the Helmholtz modes, in our case several MHz compared to several kHz) the forces on the plates due to change in liquid pressure and elasticity of the quartz have to balance, i.e. $k_p x_{1,2} = Ap_{1,2}$, where $k_p$ is the effective stiffness of the quartz plates. Substituting in the compressibility and \eqref{eq:delta-mass},\eqref{eq:delta-rho} yields the pressure in terms of helium displacement
\begin{subequations}
    \label{eq:delta-p}
    \begin{align}
        \delta p_1 &= k_p\frac{2a\rho_{sb} y_1 - a_c\rho_{sc} y_c}{\rho(2A^2 + V_B\chi k_p)}\\
        \delta p_2 &= k_p\frac{2a\rho_{sb} y_1 + a_c\rho_{sc} y_c}{\rho(2A^2 + V_B\chi k_p)}.
    \end{align}
\end{subequations}

This allows us to write the equations of motion for the three channels (neglecting damping and forcing) as
\begin{subequations}
    \label{eq:eoms}
    \begin{align}
        2\rho_{sb} al\ddot y_1 + 2\frac{\rho_{sb}}{\rho}a \delta p_1 &= 0\\
        2\rho_{sb} al\ddot y_2 + 2\frac{\rho_{sb}}{\rho}a \delta p_2 &= 0\\
        \rho_{sc}a_c l_c\ddot y_c - \frac{\rho_{sc}}{\rho}a_c(\delta p_1 - \delta p_2) &= 0,
    \end{align}
\end{subequations}
where $l, l_c$ are the lengths of the inlets and central channel, respectively. Substituting in \eqref{eq:delta-p} and transforming into Fourier space yields a linear equation $\mathbb{M}(\omega)\tilde{\bv y} = 0$ for $\bv y = (y_1, y_2, y_c)$ ($\tilde y$ denotes the Fourier transform of $y$) and the matrix $\mathbb M$ is formed by coefficients of $\tilde{y}$ in \eqref{eq:eoms}.

Finally, the Helmholtz mode frequencies are found as a condition for non-zero solutions to $\mathbb M(\omega)\tilde{\bv y} = 0$, i.e., $\det\left[\mathbb M(\omega)\right] = 0$ which yields for the fundamental mode
\begin{equation}
    \label{eq:freq-first}
    \omega_0 = \sqrt{\frac{2wD_0k_p}{2A^2l + AD\chi k_p}}\sqrt{\frac{\rho_{sb}}{\rho^2}},
\end{equation}
and for the second mode
\begin{equation}
 \label{eq:freq-second}
 \omega_1 = \sqrt{\frac{2wD_0k_p}{2A^2l + AD\chi k_p}}\sqrt{\frac{\rho_{sb}}{\rho^2} + \frac{w_cD_1l}{wD_0l_c}\frac{\rho_{sc}}{\rho^2}}.
\end{equation}

Note that apart from geometric factors, known material parameters, and the total density of liquid \4He, the frequency of the fundamental mode is determined solely by the superfluid density of the bulk liquid while the second mode is sensitive to superfluid density of both the bulk-like and strongly confined regions. Assuming that at $T=0$~K $\rho_{sc}=\rho_{sb}=\rho$, Eqs. (1) and (2) of the main text follow from Eqs.~\eqref{eq:freq-first}, \eqref{eq:freq-second}.

\section{Excitation and detection of the Helmholtz modes.}

\begin{figure}
    \centering
    \includegraphics{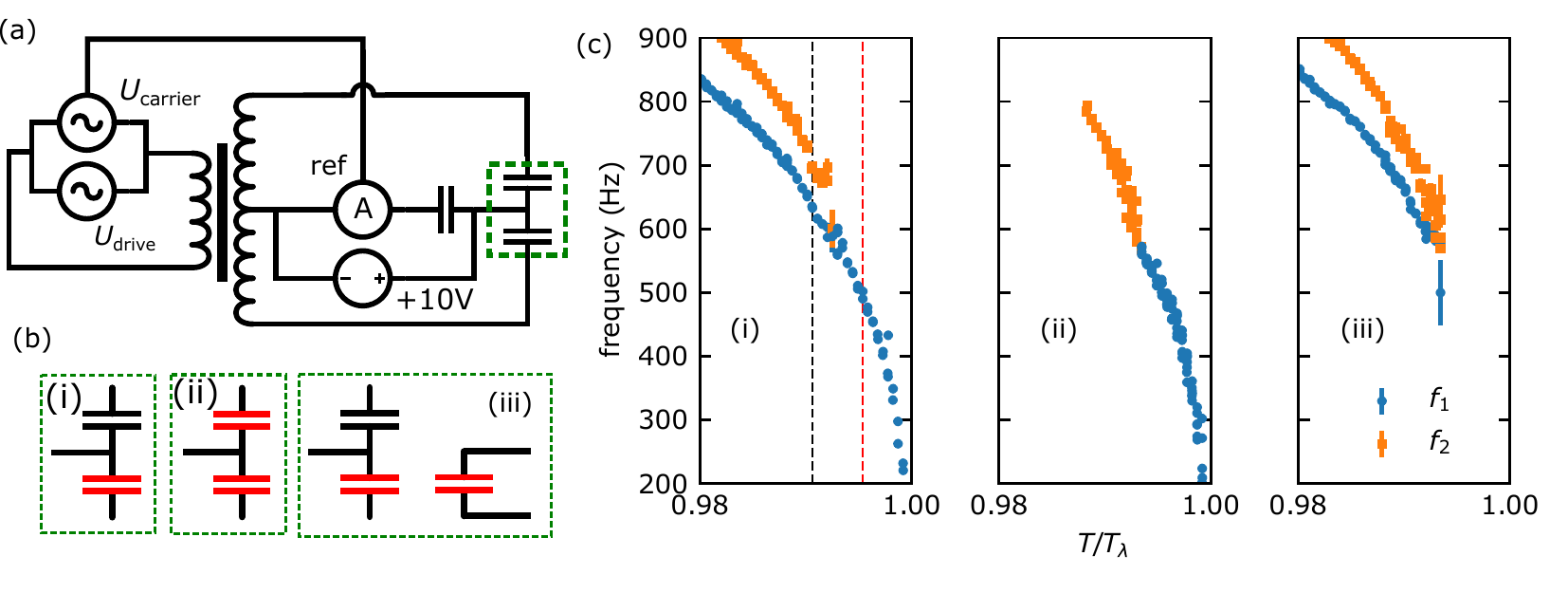}
    \caption{(a) The drive and detection scheme. The basin capacitors of the resonators are wired in a bridge circuit (in the green dashed box) in one of the three configurations shown in (b). The helium motion is detected by observing sidebands on a high-frequency (55 kHz) carrier tone ($U_\mathrm{carrier}$, see \cite{Varga2021} for details). The modes are driven by electrostatic force in the basin capacitors due to a separate tone $U_\mathrm{drive}$. To ensure that the frequency of the electrostatic force is equal to the frequency of the drive voltage the driven basins are additionally biased by +10 VDC. (b) The wiring configuration in the three measurement protocols: one-basin (i), anti-symmetric (ii) and cross-talk (iii). A red capacitor indicates a device basin, black a passive reference capacitor. In the cross-talk, the drive voltage $U_\mathrm{drive}$ and DC bias are applied through a separate circuit (not shown). (c) Temperature dependence (normalized to the bulk transition temperature $T_\lambda$) of the mode frequencies observed with the three protocols. The frequency of the fundamental mode is shown in blue and the second mode in orange (same color scheme in Fig.~\ref{fig:peaks}). The transition temperature $T_c$ of the strongly confined channel occurs at the extinction of the higher-frequency mode. The dashed vertical lines in (i) indicate the two temperatures $T<T_c$ and $T>T_c$ for which the resonant response is shown in Fig.~\ref{fig:peaks}.}
    \label{fig:protocols}
\end{figure}

\begin{figure}
    \centering
    \includegraphics{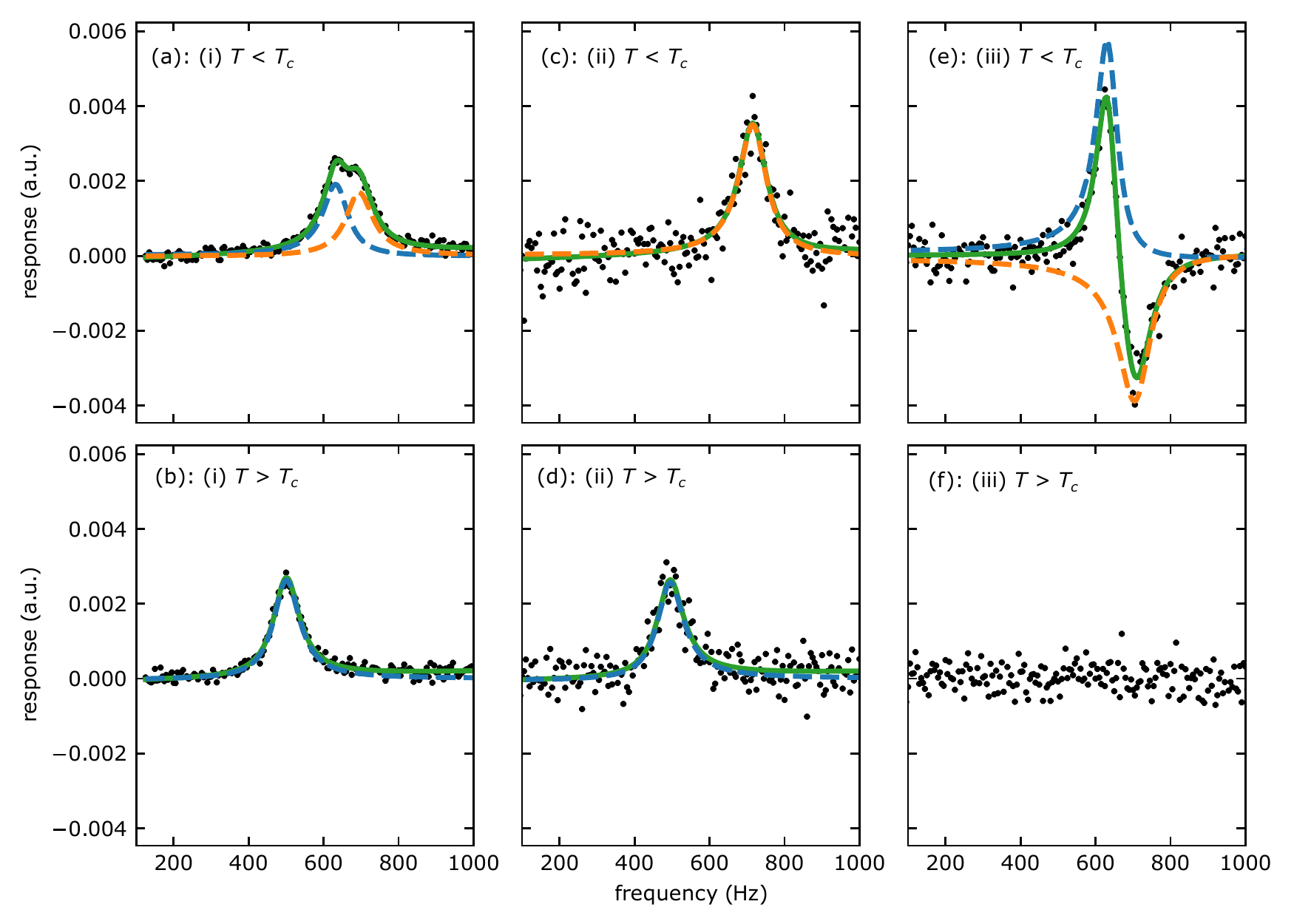}
    \caption{The resonant response of the Helmholtz device close to the transition temperature of the strongly confined channel $T_c$. (a,b) One-basin protocol. (c,d) Anti-symmetric protocol. (e,f) Cross-talk protocol. The scatter plots show the quadrature of the response measured at the two temperatures indicated by vertical lines in Fig.~\ref{fig:protocols}(c) (same for all protocols). Also shown are fits to a linear harmonic oscillator response. Full green lines -- full fit to one or two peaks (depending on protocol and temperature) and background. Blue dashed line -- peak corresponding to the fundamental mode; dashed orange line -- peak corresponding to the second Helmholtz mode (cf. Fig.~\ref{fig:protocols}(c)).}
    \label{fig:peaks}
\end{figure}

The electrical scheme for forcing and readout of the Helmholtz modes is shown in Fig.~\ref{fig:protocols}(a,b). The basin capacitors are wired in a tuned bridge circuit and the motion is detected as the bridge imbalance. The driven basins are biased by a 10 VDC source and Helmholtz mode is driven by voltage $U_\mathrm{drive}$. The motion is sensed by observing sidebands induced on a carrier tone $U_\mathrm{carrier}$ (see \cite{Varga2021} for details). Fig.~\ref{fig:protocols}(b) shows the three different measurement protocols (i) -- (iii), which differ by configuration of the two basin capacitors for either drive or detection. The sensitivity of the three protocols to the two modes is shown in Fig.~\ref{fig:protocols}(c), which shows the frequencies of the peaks observed in the response obtained by the protocols (i)--(iii). The observed peaks are shown in Fig.~\ref{fig:peaks} for the two temperatures indicated in Fig.~\ref{fig:protocols}(c,(i)). The one-basin protocol (i) overlaps in forcing and detection with both modes and hence both modes are driven and observed (Fig.~\ref{fig:peaks}(a)). Similarly the cross-talk (iii) is also sensitive to both modes, however, once the basins become decoupled the response disappears. Finally, the anti-symmetric protocol (ii) can only drive and detect the mode when pressure oscillates in the two basins with a 180\textdegree~phase shift. Below $T_c$, this is the second Helmholtz mode. Above $T_c$ when the basins are decoupled, the fundamental mode is degenerate with respect to the relative phase (the basins form two independent resonators) and hence the fundamental mode is observed (Fig.~\ref{fig:peaks}(c,d)).

\section{Characterization of the strongly confined channel}

\begin{figure}
    \centering
    \includegraphics{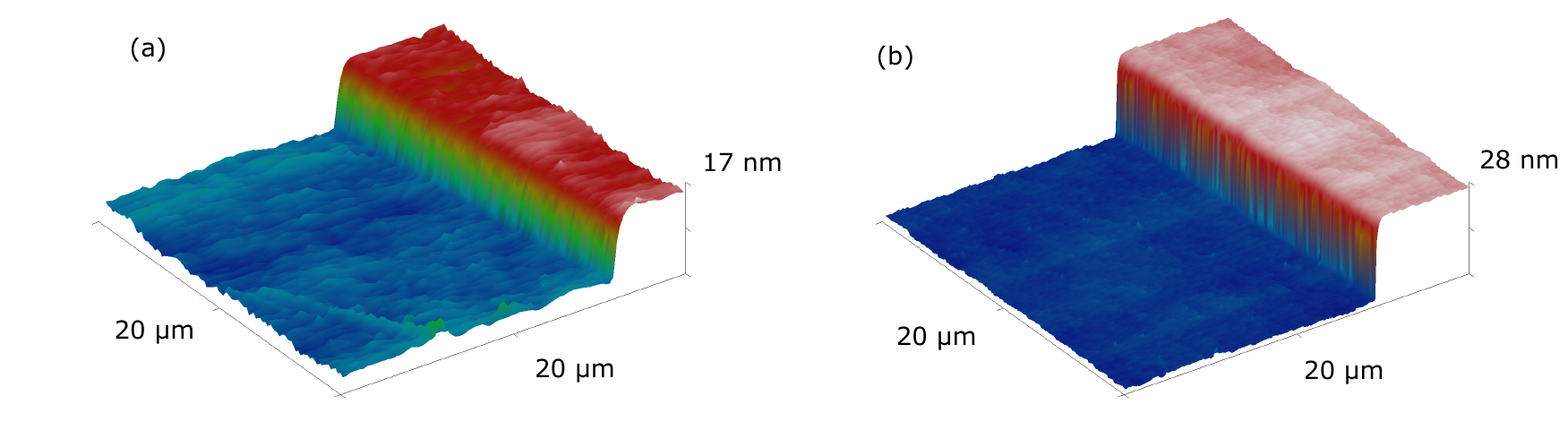}
    \caption{AFM imaging of the edge of the strongly confined channel. (a) 25 nm device (average step height 12.3 nm). (b) 50 nm device (average step height 24.6 nm). Shown is only one side of the valley formed by wet etching of the quartz chips. The actual device channel is formed when two identical chips are bonded together in the direction perpendicular to the device plane. That is, the actual height of the channel is twice the measured step height.}
    \label{fig:afm}
\end{figure}

Atomic force microscopy imaging (AFM) of the edge of the strongly confined central channel is shown in Fig.~\ref{fig:afm}. The imaged resonators are not same as the ones used for measurement. They were, however, fabricated from the same substrate wafer and in the same batch by wet etching in the etchant bath for identical length of time. It is only possible to image the valley formed by etching in an un-bonded device, i.e., the channel in the actual device is twice as high. Based on AFM imaging on multiple locations along the channel edge we find that for the 25 nm device the average channel size is $24.6\pm 0.2$~nm with surface roughness inside the channel of $0.7\pm0.1$~nm rms and for the 50 nm device the channel size $49 \pm 1$~nm with surface roughness of $0.6\pm0.1$~nm rms.

\section{Bulk and surface contributions to the normal density}

\begin{figure}
    \centering
    \includegraphics{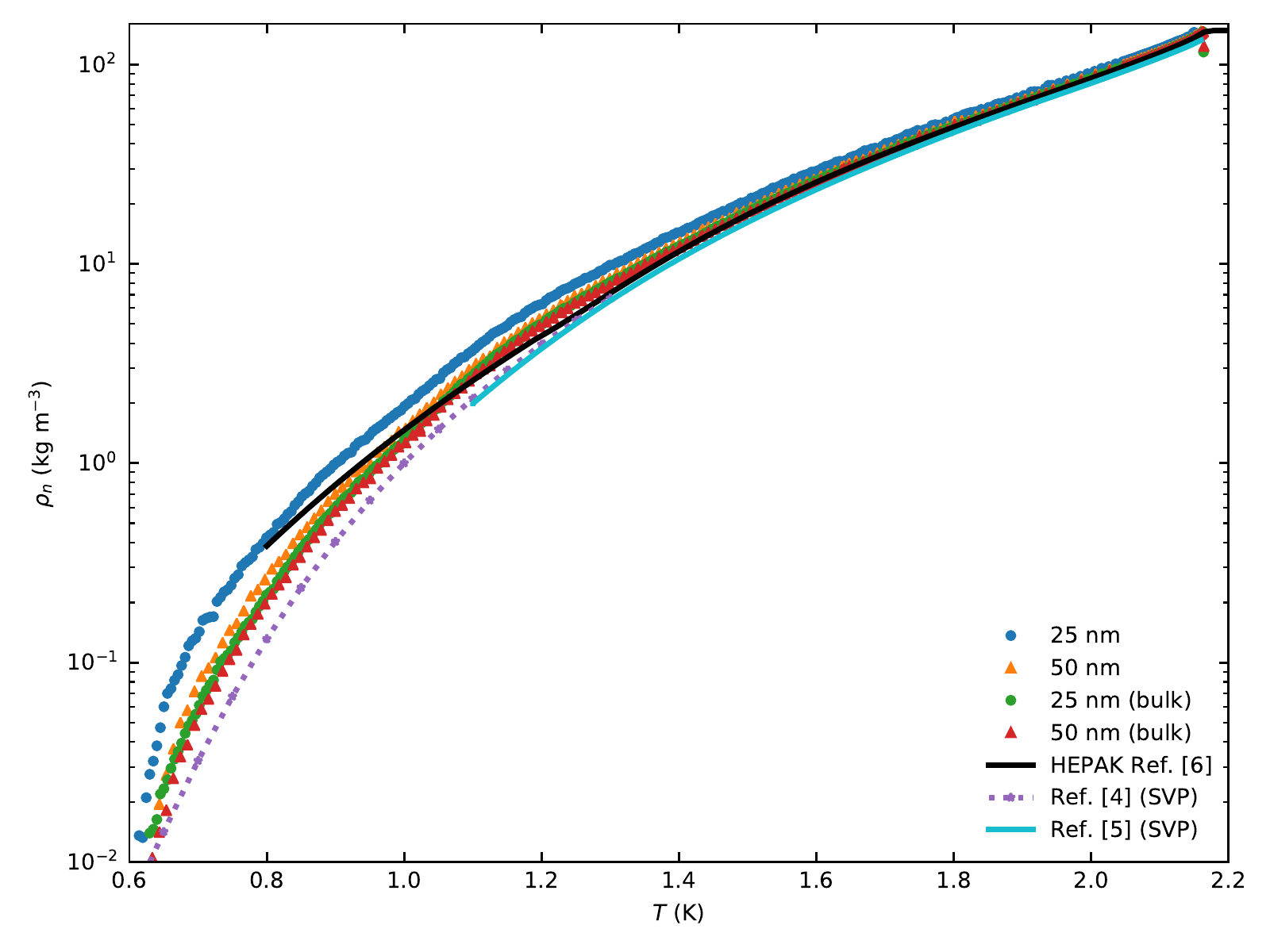}
    \caption{Bulk and confined normal fluid densities for both 25 nm and 50 nm devices as a function of temperature measured at approximately 1 bar. The slight systematic mismatch between the two bulk values is due to slightly different pressure (approximately 1073 and 1067 mbar, respectively, for 50 nm and 25 nm). The obtained data are in good qualitative and quantitative agreement with database values \cite{Donnelly1998} (SVP) and \cite{HEPAK} (1067 mbar) and calculation based on neutron scattering data \cite{Godfrin2021}. The sharp bend just above 0.6 K is due to the 0.6 K normalization of \eqref{eq:freq-first} and \eqref{eq:freq-second}.}
    \label{fig:rhon-raw}
\end{figure}

In order to analyze the contribution of the surface excitations to the observed normal fluid density in the confined channel it is necessary to carefully subtract the contribution of the bulk.  Thanks to the two-basin design of our device, both bulk and confined normal densities are measured \emph{independently and in-situ}, which removes ambiguity in the estimation of the bulk normal fluid density. The two independently measured densities are shown in Fig.~\ref{fig:rhon-raw} for both 25 nm and 50 nm devices at approximately one bar. We find acceptable agreement for the bulk normal density with database values and recent experiments, although, exhaustive and authoritative determination of the absolute normal fluid density is beyond the scope of this work. The total normal fluid density results from the entire dispersion curve of helium and separation into phonon and roton contributions is generally inaccurate \cite{Godfrin2021}. 

The subtraction of the two independently measured densities leads to the normal fluid density enhancement as shown in Fig.~3 of the main text. The added normal fluid density is fitted with expression (Eq. 6 of the main text)
\begin{equation}
\label{eq:sm-roton-rhon}
    \Delta\rho_n(T) = \frac{2}{D}\frac{\hbar k_0^3 m^{*1/2}}{2\sqrt{2\pi k_B T}}\exp\left(-\frac{\Delta}{k_B T}\right) = \frac{2}{D}A_R T^{-1/2}\exp\left(-\frac{\Delta}{k_B T}\right).
\end{equation}

\begin{figure}
    \centering
    \includegraphics{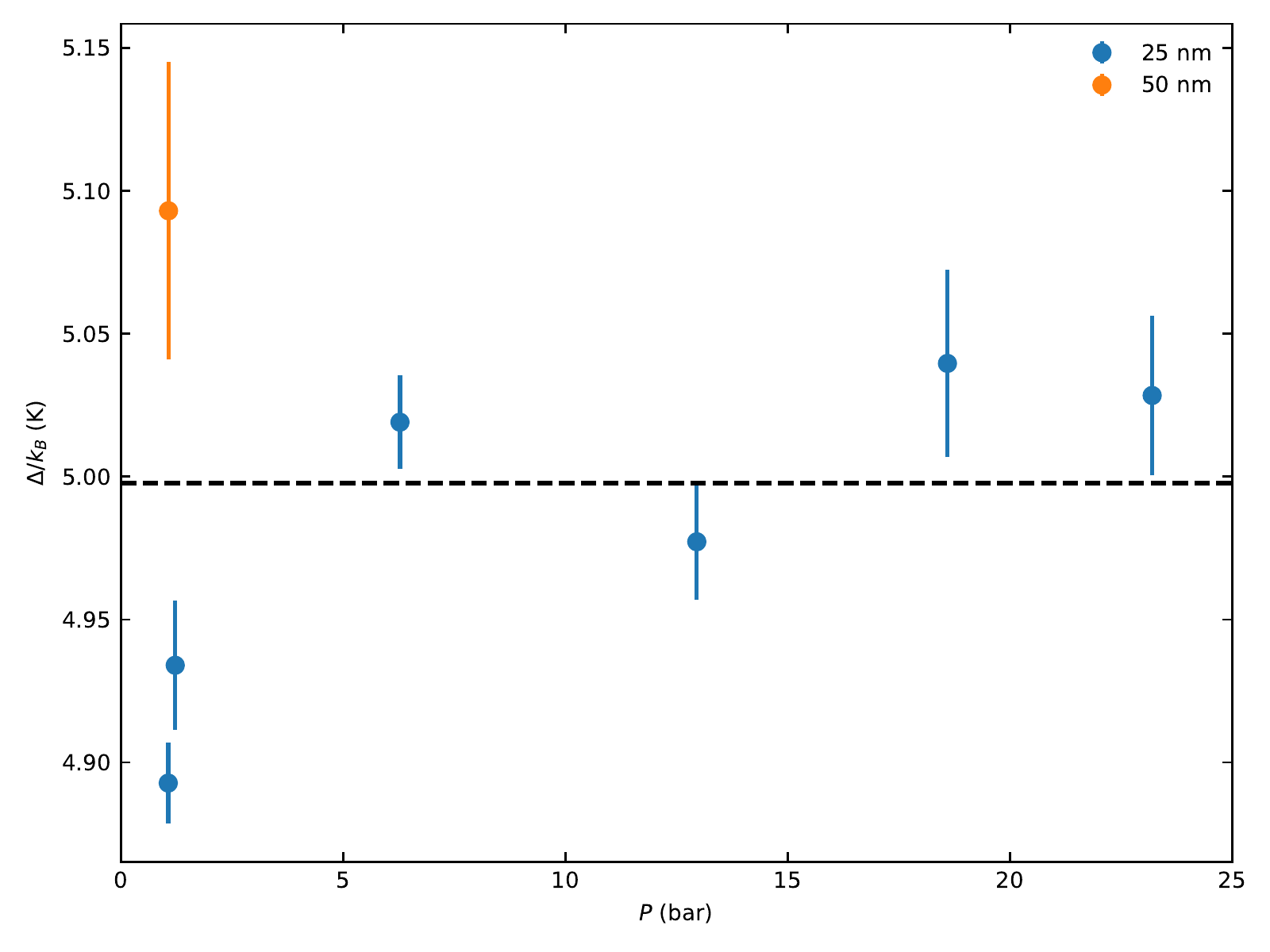}
    \caption{Values of the roton gap for all pressures and both confinements. The error bars were estimated    using bootstrapping \cite{Efron1981}. The horizontal line shows the mean $\bar{\Delta}/k_B$ = 4.99$\pm$0.06 K.}
    \label{fig:deltas}
\end{figure}

The observed roton gaps, deduced from a linear fit in the region $T\in[0.8, 1.5]$~K in Fig.~3(a,b) of the main text are shown in Fig.~\ref{fig:deltas}. These data were obtained from three separate cooldowns of the refrigerator: one cooldown for the 50 nm device and two separate cooldowns for the 25 nm, one for 1 bar data, one for the pressure dependence; devices were exposed to air at room temperature in-between cooldowns. Note that \eqref{eq:sm-roton-rhon} describes the data well in the entire temperature range, however the fitting range was reduced due to increased scatter at high temperatures (due to less accurate determination of mode frequencies) and deviation at low temperatures due to the fact that equations (2) and (3) of the main text were normalized by frequencies at 0.6 K rather than by true 0 K limit frequencies. With the exception of these regions the fit values are independent of fitting range (reducing the fit range to 0.9 -- 1.4 K shifts down the average $\Delta$ by less than 1\% and slightly increases the scatter of the points).

The prefactor $A_R$ in \eqref{eq:sm-roton-rhon} contains information about the roton wave vector and effective mass, which cannot be separated from the fitting of the data alone. These quantities are, in principle, available from simulations. However, reliable estimates of the effective mass are generally difficult to obtain \cite{Krotscheck2015} and the 2D density of the film, typically used as an input in the simulations \cite{Krotscheck2015,Arrigoni2013}, is not known experimentally in our case (i.e., we relate the measured superfluid fraction to the bulk density).

With these caveats in mind, we can estimate the roton wave vector by assuming, since $A_R$ varies more slowly with $m^*$, a constant effective roton mass $m^*\approx 0.2m_4$ \cite{Padmore1974,Chester1976}. The resulting wave vectors, shown in Fig.~\ref{fig:wave-vectors}, lie in the range 1.6~\AA -- 2.3~\AA. Note that the value of the wave vector depends on the value of the assumed effective mass as $k_0 \propto m^{*-1/6}$. Considering the crude estimate of the effective mass, the value of the wave vector is in good quantitative agreement with both numerical results \cite{Apaja2003a,Krotscheck2015,Arrigoni2013} and neutron scattering data \cite{Clements1996,Bossy2012,Bossy2019}. 

\begin{figure}
    \centering
    \includegraphics{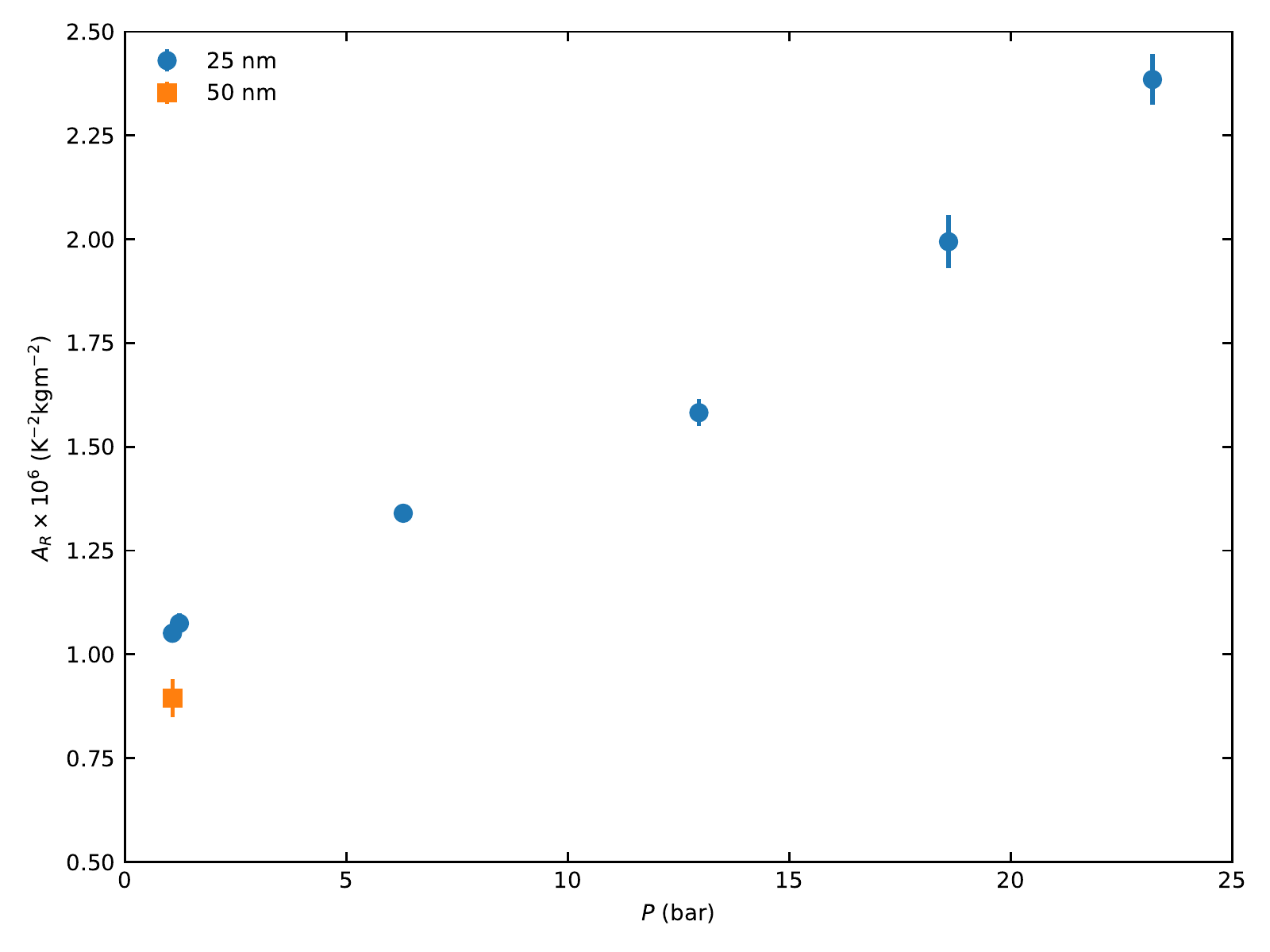}
    \caption{Pressure dependence of the prefactor $A_R$ of the fit of \eqref{eq:roton-rhon} to the added normal fluid density for both 25 and 50 nm confined channels.}
    \label{fig:prefactors}
\end{figure}

\begin{figure}
    \centering
    \includegraphics{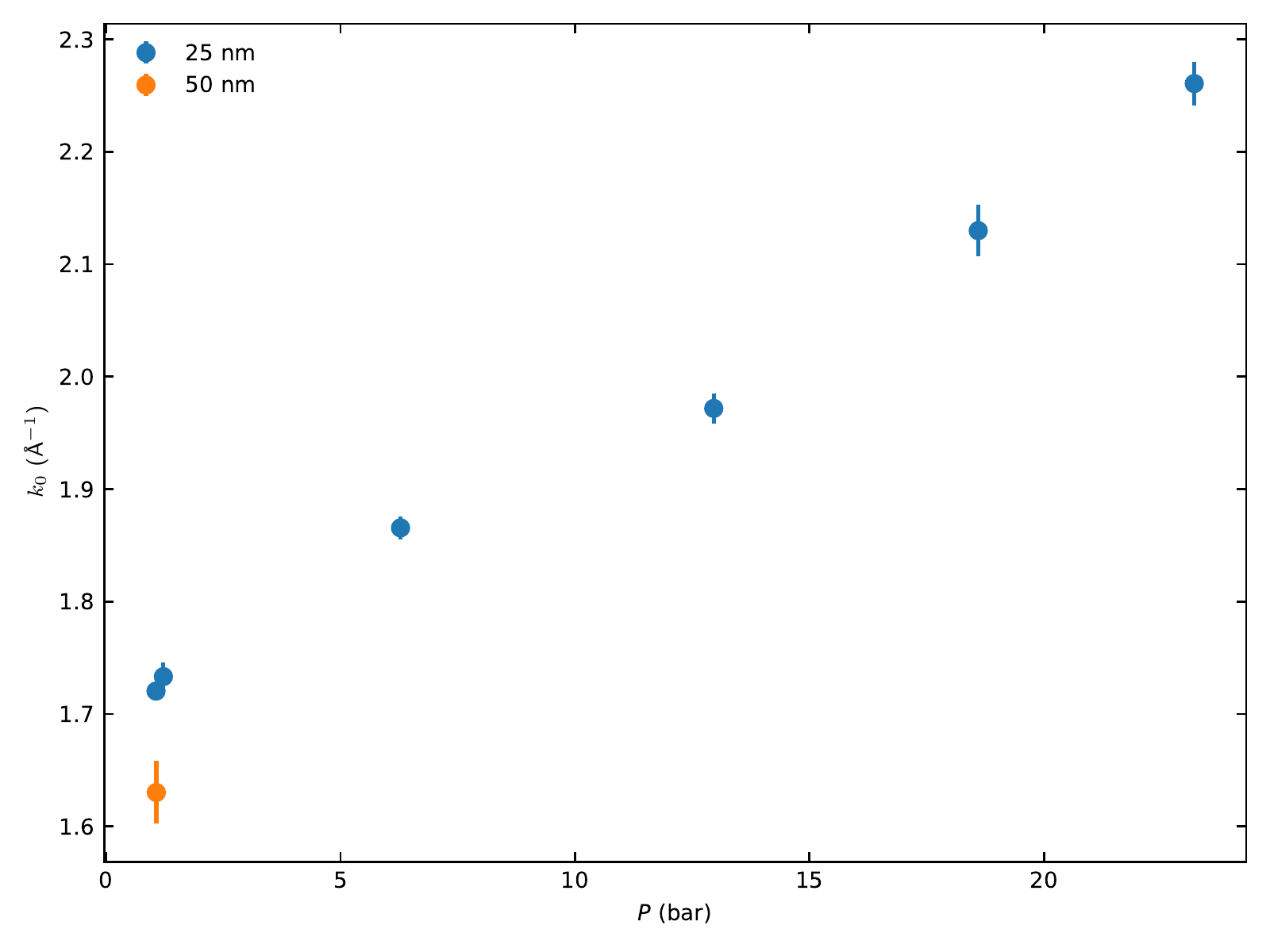}
    \caption{Two-dimensional roton wave vectors calculated from the prefactor $A_R$ of the roton contribution to the normal fluid density \eqref{eq:sm-roton-rhon} assuming fixed effective roton mass $m^*=0.2m_4$.}
    \label{fig:wave-vectors}
\end{figure}

\bibliography{confined-helium}

\end{document}